\documentclass[lettersize,journal]{IEEEtran}
\usepackage{amsmath,amsfonts}
\usepackage{algorithmic}
\usepackage{algorithm}
\usepackage{array}
\usepackage[caption=false,font=normalsize,labelfont=sf,textfont=sf]{subfig}
\usepackage{textcomp}
\usepackage{stfloats}
\usepackage{url}
\usepackage{verbatim}
\usepackage{graphicx}
\usepackage{cite}
\usepackage{upgreek}

\begin{document}
\bstctlcite{IEEEexample:BSTcontrol}

\title{\huge A 16-Channel Low-Power Neural Connectivity Extraction\\ and Phase-Locked Deep Brain Stimulation SoC}

\author{Uisub~Shin,~\IEEEmembership{Student Member,~IEEE,}~Cong~Ding,~\IEEEmembership{Student Member,~IEEE,}\\~Virginia~Woods,~Alik~S.~Widge,~and~Mahsa~Shoaran,~\IEEEmembership{Member,~IEEE\vspace{-10mm}}

\thanks{Manuscript received July 1, 2022. This work was supported by the National Institute of Mental Health Grant R01-MH-123634. All animal experimental procedures were approved by the Institutional Animal Care and Use Committees (IACUC) at the University of Minnesota under IACUC Protocol No. 2105-39119A. \textit{(Corresponding author: Uisub Shin.) }}%
\thanks{Uisub Shin is with the Institute of Electrical and Micro Engineering, EPFL, 1202 Geneva, Switzerland, and the School of Electrical and Computer Engineering, Cornell University, Ithaca, NY 14853 (e-mail: us52@cornell.edu).}
\thanks{Cong Ding and Mahsa Shoaran are  with the Institute of Electrical and Micro Engineering and Center for Neuroprosthetics, EPFL, 1202 Geneva, Switzerland.}
\thanks{Virginia Woods and Alik S. Widge are with the Department of Psychiatry and Behavioral Sciences, University of Minnesota, Minneapolis, MN 55455.}}

\markboth{$>$ REPLACE THIS LINE WITH YOUR MANUSCRIPT ID NUMBER (DOUBLE-CLICK HERE TO EDIT) $<$}%
{Shell \MakeLowercase{\textit{et al.}}: A Sample Article Using IEEEtran.cls for IEEE Journals}

\maketitle

\begin{abstract}
Growing evidence suggests that phase-locked deep brain stimulation (DBS) can effectively regulate abnormal brain connectivity in neurological and psychiatric disorders. This letter therefore presents a low-power SoC with both neural connectivity extraction and phase-locked DBS capabilities. A 16-channel low-noise analog front-end (AFE) records local field potentials (LFPs) from multiple brain regions with precise gain matching. A novel low-complexity phase estimator and neural connectivity processor subsequently enable energy-efficient, yet accurate measurement of the instantaneous phase and cross-regional synchrony measures. Through flexible combination of neural biomarkers such as  phase synchrony and spectral energy, a four-channel charge-balanced neurostimulator is triggered to treat various pathological brain conditions. Fabricated in 65nm CMOS, the SoC occupies a silicon area of 2.24mm\textsuperscript{2} and consumes 60$\boldsymbol\upmu$W, achieving over 60\% power saving in neural connectivity extraction compared to the state-of-the-art. Extensive \emph{in-vivo} measurements demonstrate multi-channel LFP recording, real-time extraction of phase and neural connectivity measures, and phase-locked stimulation in rats.
\end{abstract}


\section{Introduction}
Deep brain stimulation (DBS) is a well-established therapy for movement disorders such as Parkinson's disease (PD) and essential tremor. However, the conventional open-loop  DBS causes stimulation-induced side effects, can consume high energy \cite{cagnan2017stimulating, yao2020improved}, and lacks sufficient efficacy in emerging applications such as psychiatric and memory disorders  \cite{sullivan2021deep}. It may be more efficient and effective to deliver DBS locked to the phase of ongoing neural oscillations \cite{sullivan2021deep, wendt2022physiologically}. This approach has already shown promise in movement disorders~\cite{cagnan2017stimulating}. Excessive inter-regional connectivity indicates pathological brain states in PD~\cite{de2013exaggerated} and network-based diseases such as depression~\cite{sullivan2021deep}. Continuous monitoring of neural connectivity and its regulation via phase-locked DBS could provide a new solution to treating such  disorders. 

The CORDIC processors in \cite{abdelhalim201364, o2018recursive} can accurately compute the instantaneous phase and neural connectivity measures such as phase locking value (PLV) and phase-amplitude coupling (PAC). However, the high accuracy comes at the cost of significant power consumption ($>$200$\upmu$W). Furthermore, the clock latency in the pipelined CORDIC would hinder real-time phase-locked stimulation. Ref. \cite{delgado2019phase} reported a low-complexity PLV approximation algorithm based on local minima detection. While efficient, this approach compromises PLV accuracy and cannot track instantaneous phase and amplitude in real time for oscillation-locked stimulation.

This letter extends upon our first-in-literature phase-locked DBS SoC \cite{shin202216} and presents circuit details, new analyses, and measurement results. An energy-efficient neural connectivity processor is proposed to overcome the energy-accuracy bottleneck in the existing designs. The phase-locked DBS prototype is extensively validated via benchtop and \emph{in-vivo} testing.

\begin{figure}[t]
\centering
\includegraphics[width=1\columnwidth]{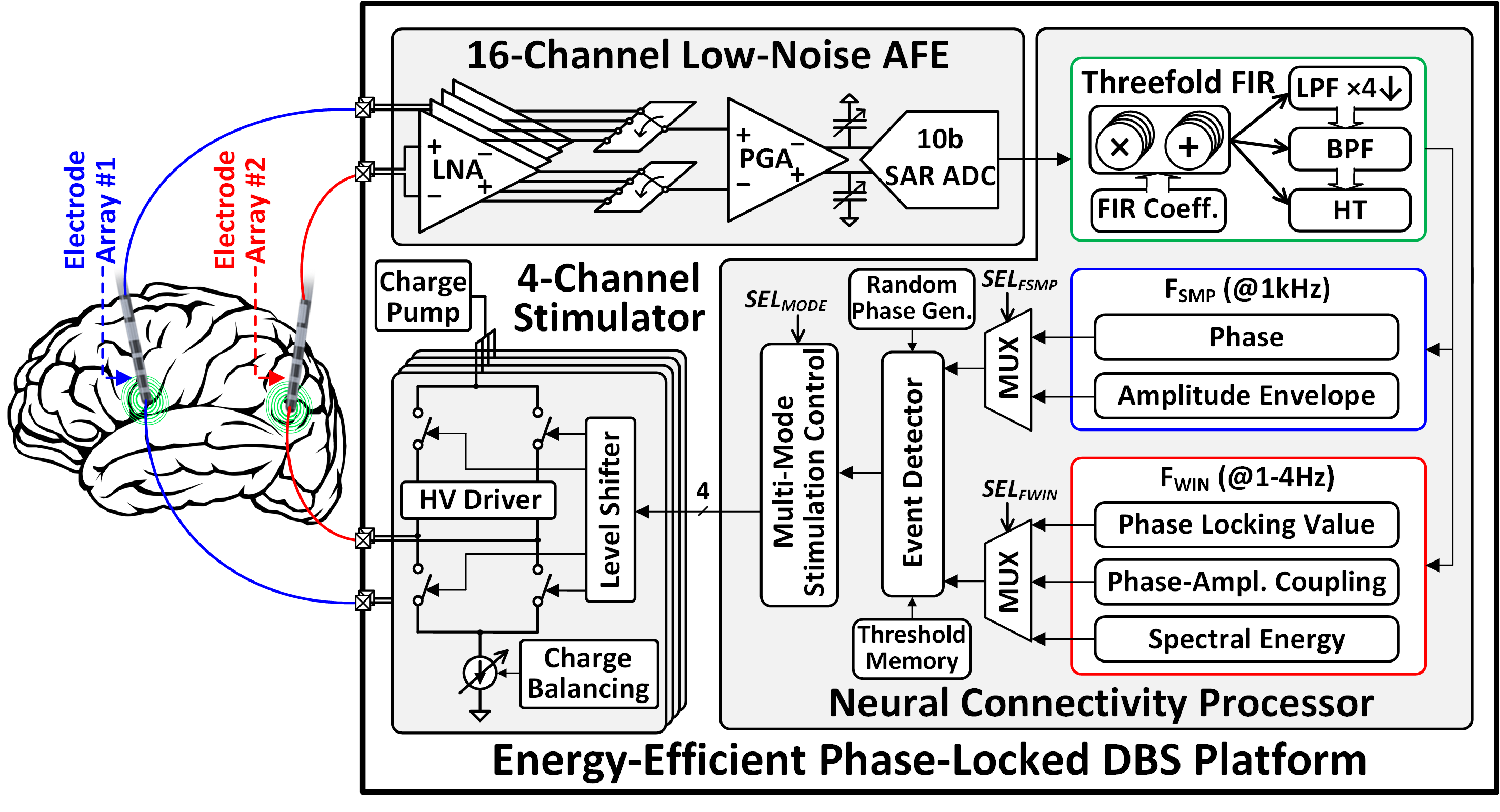}
\vspace{-6mm}
\caption{Architecture of the proposed 16-channel neural connectivity extraction and phase-locked DBS SoC.}
\vspace{-3mm}
\label{f1_SoC}
\end{figure}

\section{SoC Architecture}
To effectively regulate cross-regional  connectivity, the system must have: 1) low-noise multi-channel local field potential (LFP) recording with precise gain matching, 2) accurate, continuous extraction of  instantaneous phase and connectivity measures,  and 3) charge-balanced stimulation to minimize tissue damage.

Fig. \ref{f1_SoC} presents the architecture of the proposed phase-locked DBS SoC that meets these criteria. A 16-channel chopper-stabilized  analog front-end (AFE) conditions LFPs from at least two electrode arrays (e.g., DBS leads). The 16-channel digitized LFPs are sent to the proposed neural connectivity processor for phase-locked DBS control. A programmable finite impulse response (FIR) filter performs lowpass filtering (LPF) for decimation, bandpass filtering (BPF), and Hilbert transform (HT) through hardware sharing. The filtered signals are processed in a feature extractor (FE), where the instantaneous phase and amplitude envelope are extracted per sample, and PLV, PAC, and spectral energy (SE) on a window-by-window basis. Upon detection of pathological neural activity (e.g., excessive PLV or PAC via feature thresholding), a four-channel high-voltage compliant stimulator delivers  charge-balanced biphasic  pulses to the brain, timed to a specific  phase of neural oscillations.

\begin{figure}[t]
\centering
\includegraphics[width=1\columnwidth]{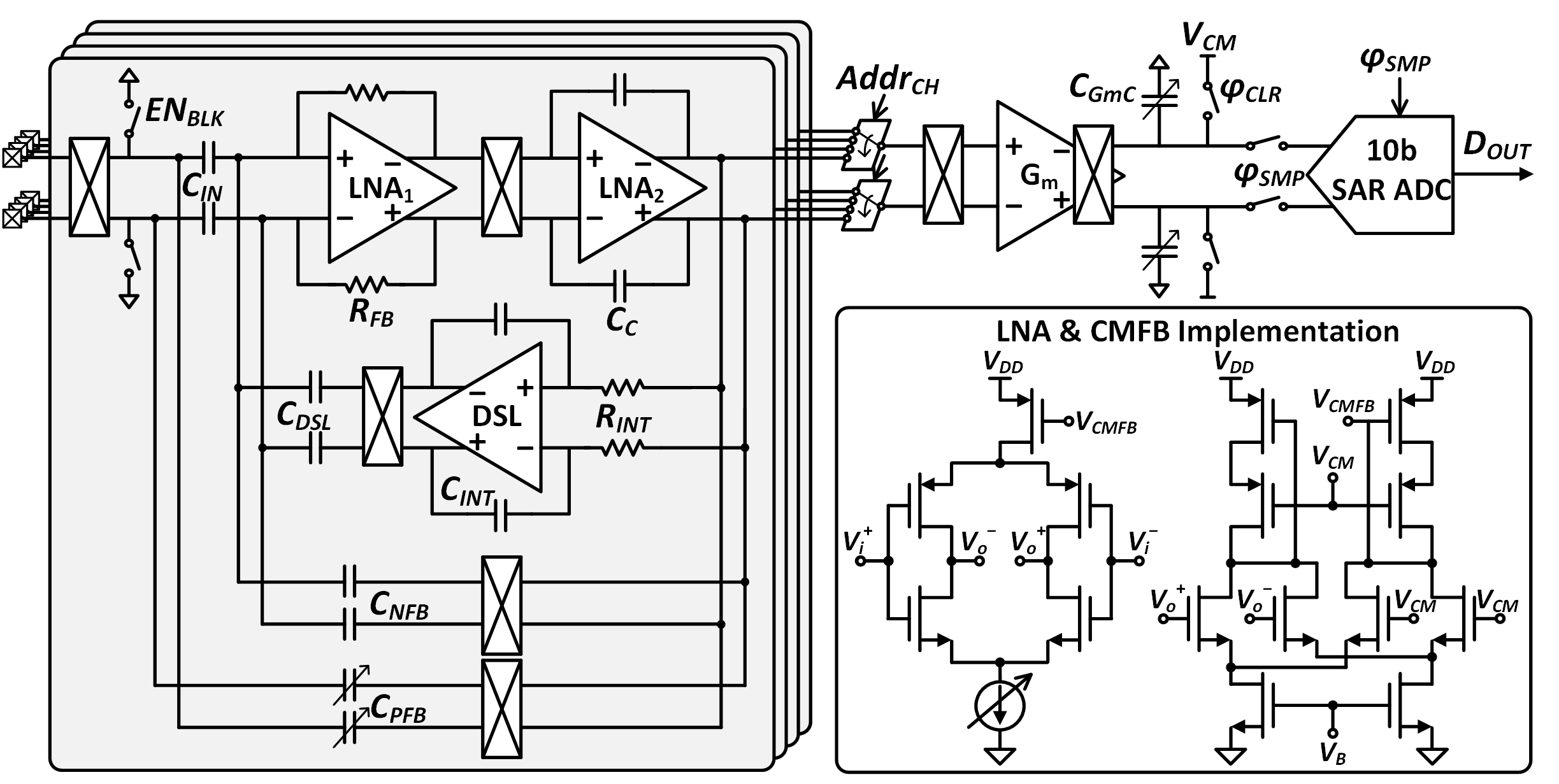}
\vspace{-6mm}
\caption{Circuit implementation of the 16-channel chopper-stabilized AFE.}
\vspace{-3mm}
\label{f2_AFE}
\end{figure}

\begin{figure}[t]
\centering
\includegraphics[width=1\columnwidth]{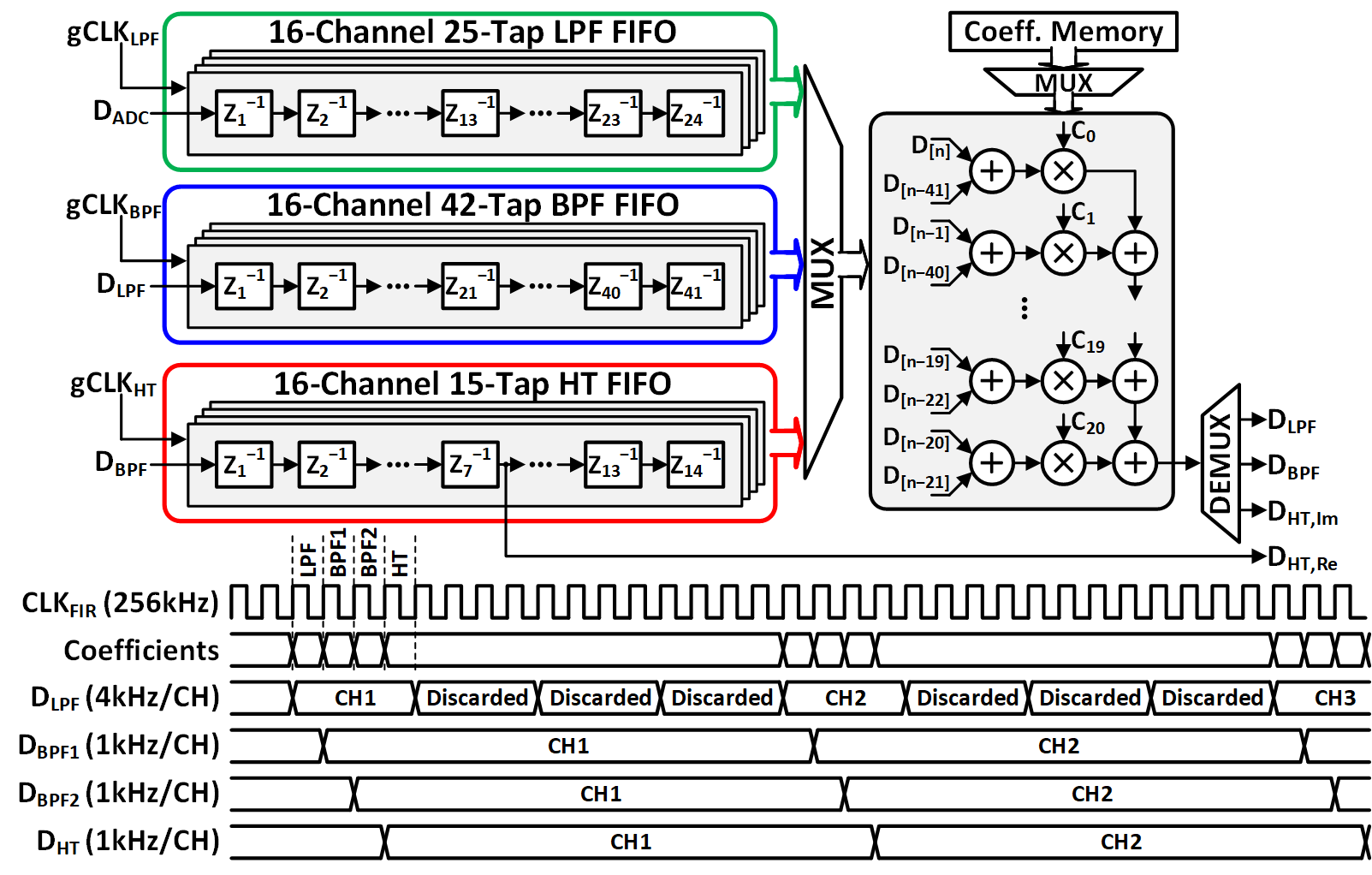}
\vspace{-6mm}
\caption{Programmable threefold FIR filter architecture and timing diagram.}
\vspace{-3mm}
\label{f3_FIR}
\end{figure}

\section{Circuit Implementation}
\subsection{Low-Noise Analog Front-End}
Fig. \ref{f2_AFE} presents the circuit implementation of the 16-channel AFE. Two-stage low-noise amplifiers (LNAs) are implemented with a current-reuse inverter topology for enhanced noise efficiency. The noise performance is further improved by chopper stabilization that suppresses the flicker noise in the LFP band (1--500Hz). The DC servo loop (DSL) cancels electrode DC offsets, and the tunable positive feedback capacitor boosts the input impedance. To save chip area, the 16-channel LNA outputs are multiplexed into a programmable-gain integrator followed by a 10-bit successive approximation register (SAR) analog-to-digital converter (ADC). The 16-channel closed-loop LNAs with the shared integrator achieve precise gain matching, which is crucial for extracting cross-regional connectivity measures. A 16:1 multiplexer can address the input channels in any user-defined order, allowing extraction of a flexible combination of biomarkers in the digital processor. When stimulation is triggered, blanking is performed by turning off the input chopper and resetting the AFE input to prevent amplifier saturation due to stimulation artifacts.

\begin{figure}[t]
\centering
\includegraphics[width=1\columnwidth]{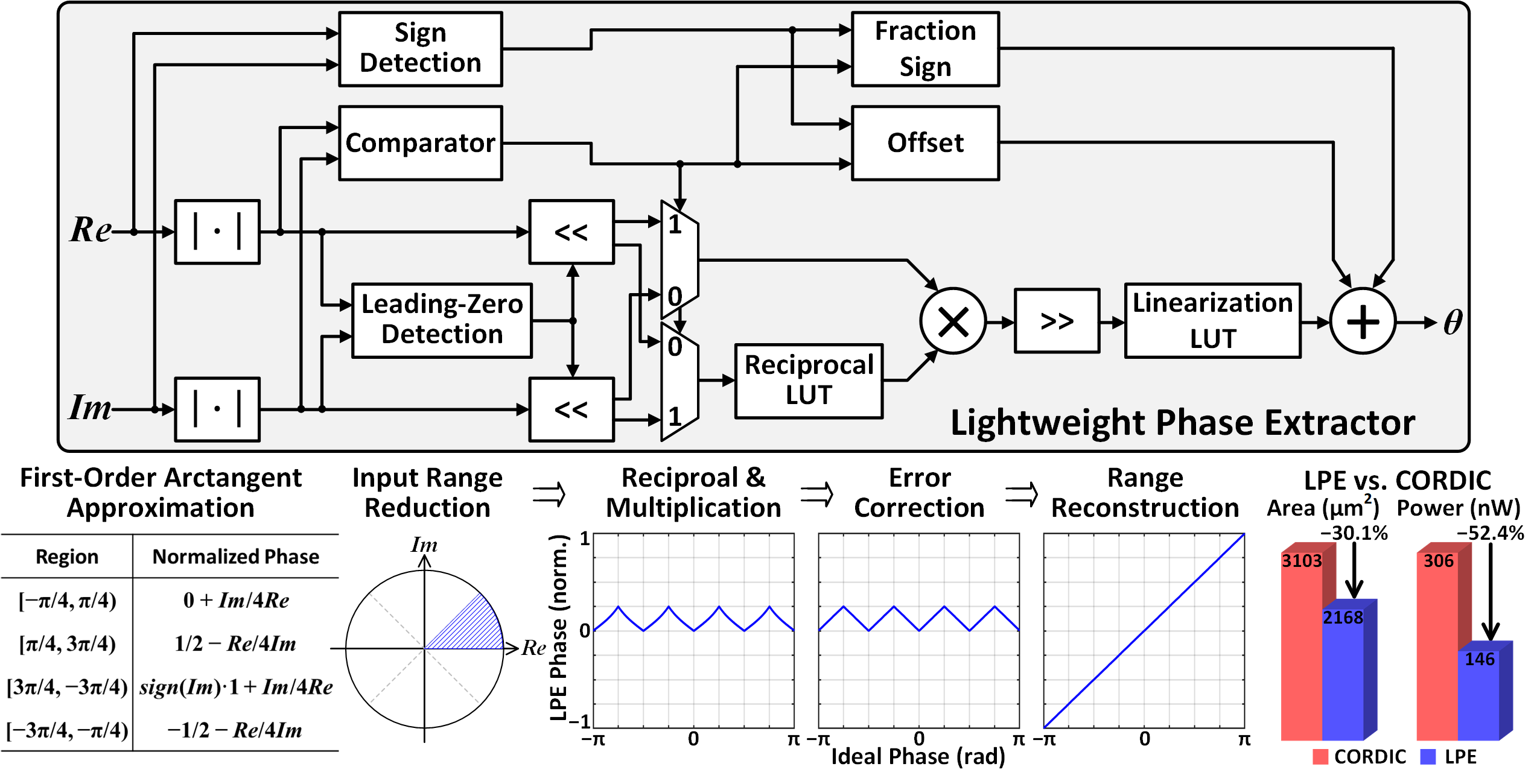}
\vspace{-6mm}
\caption{Proposed LPE hardware and  illustration of the underlying algorithm.}
\vspace{-3mm}
\label{f4_LPE}
\end{figure}

\begin{figure}[t]
\centering
\includegraphics[width=1\columnwidth]{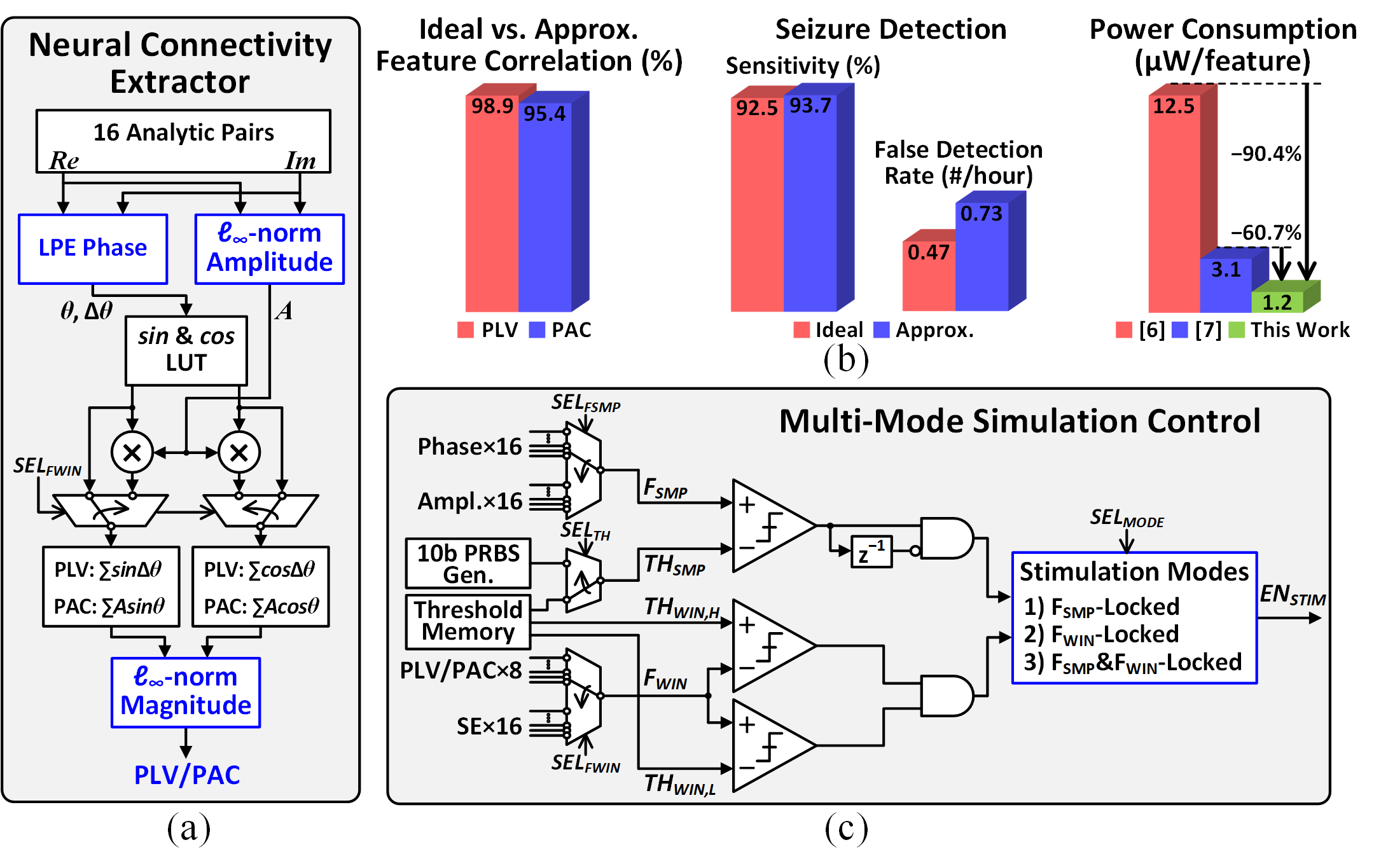}
\vspace{-6mm}
\caption{(a) Proposed neural connectivity extractor, (b) feature accuracy analysis and power comparison, and (c) multi-mode stimulation control.}
\vspace{-3mm}
\label{f5_FE}
\end{figure}

\vspace{-3mm}
\subsection{Neural Connectivity Processor}
The programmable threefold FIR architecture is depicted in Fig. \ref{f3_FIR}. Three banks of delay registers (LPF, BPF, and HT) share a single multiplier-adder chain to reduce chip area. The FIR decimates the 16-channel AFE output by a factor of 4 and generates 16 bandpass-filtered signals and 16 analytic pairs (\textit{Re} and \textit{Im}) in any pre-programmed frequency bands. The clock for each delay line is individually gated, reducing the effective switching frequency of registers to 4kHz (LPF) and 1kHz (BPF and HT). Along with data gating, clock gating saves dynamic power by preventing unnecessary switching activity.

\begin{figure}[t]
\centering
\includegraphics[width=0.94\columnwidth]{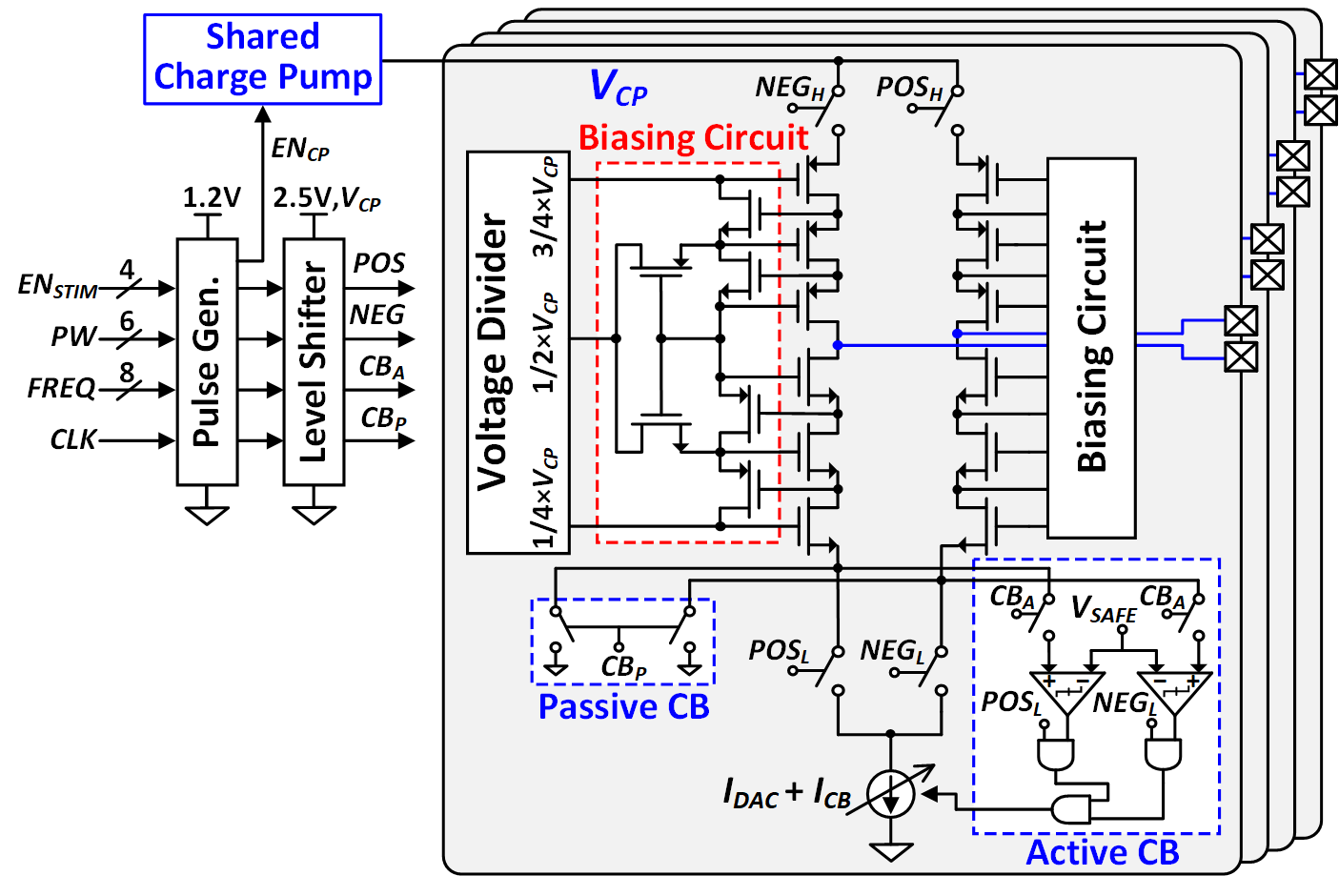}
\vspace{-2mm}
\caption{High-voltage compliant charge-balanced neurostimulator.}
\vspace{-3mm}
\label{f6_Stim}
\end{figure}

To facilitate low-complexity phase extraction on chip, we propose a Lightweight Phase Extractor (LPE) that achieves a superior power-accuracy trade-off compared to conventional methods. Fig. \ref{f4_LPE} shows the LPE hardware implementation and underlying approximation algorithm based on first-order Lagrange interpolation \cite{rajan2006efficient}. Here, sign detection and magnitude comparison are used to identify the region in the complex plane for the input analytic pair. Bit-scaling and numerator-denominator selection are subsequently performed to confine the pair to the [0, $\uppi$/4) range. The fraction of the real and imaginary signals is then calculated using a reciprocal look-up table (LUT) and a multiplier. To improve phase accuracy, the first-order approximation errors ($\simeq$4$^{\circ}$) are corrected by a linearization LUT. Thanks to the preceding input range reduction, the size of the reciprocal and linearization LUTs is reduced to 2\textsuperscript{8}$\times$9 and 2\textsuperscript{8}$\times$7 bits, respectively, for 10-bit phase extraction. Finally, the trigonometric periodicity identities are exploited to reconstruct the [--$\uppi$, $\uppi$) range by adding the corresponding offset value to the error-corrected fraction. The LPE generates normalized 10-bit phase outputs that are last-bit accurate with respect to an infinite-precision arctangent operator in MATLAB. Compared to a bit-width optimized,  unrolled CORDIC (10-bit), the LPE is 30.1\% more area efficient while consuming 52.4\% less power.

Fig. \ref{f5_FE}(a) presents the neural connectivity extractor. In addition to the instantaneous phase, the ideal PAC~\cite{canolty2006high} and PLV \cite{mormann2000mean} extraction involves the computation of  instantaneous amplitude envelope and/or magnitude. For improved hardware efficiency, the complex Euclidean norm is approximated by the \emph{$l_\infty$}-norm, 
which takes the larger of the two absolute input values as  output. We validated these approximations with a feature correlation analysis and a seizure detection task, using  an intracranial EEG  dataset (iEEG.org) and the gradient boosting model proposed in \cite{shoaran2018energy}. Fig. \ref{f5_FE}(b) shows that the approximated features are $>$95\% correlated with the ideal ones and only marginally affect the seizure detection performance. The proposed neural connectivity extractor can simultaneously extract 8 PAC/PLVs, 16 phases, and 16 amplitude envelopes from any combination of 8 channel pairs with only 9.69$\upmu$W power consumption including the FIR. This is $>$60.7\% power saving compared to the CORDIC-based designs \cite{abdelhalim201364, o2018recursive}. 

To allow flexible stimulation control for different therapeutic settings and  symptoms, the SoC supports multiple stimulation modes by tracking: 1) per-sample phase or amplitude envelope, 2) windowed PLV, PAC, or SE, or 3) a combination of the two, as depicted in Fig. \ref{f5_FE}(c). For instance, stimulation  can be locked to the phase crossing at a target value (\textit{TH\textsubscript{SMP}}), while  the neural connectivity measure (PLV or PAC) lies within a pre-defined therapeutic range set by \textit{TH\textsubscript{WIN,L}} and \textit{TH\textsubscript{WIN,H}}. As a unique feature of this SoC, randomized phase locking can also be enabled using a pseudo-random binary sequence (PRBS) threshold generator. This may be particularly useful for disrupting a target oscillation in the brain. 

\vspace{-3mm}
\subsection{Charge-Balanced Neurostimulator}
Fig. \ref{f6_Stim} depicts the four-channel charge-balanced neurostimulator with a high-voltage compliant stacked architecture. The biasing circuit maintains the bias voltage for each transistor within a tolerance range under the 8V output voltage of the charge pump. To precisely match the biphasic stimulation pulses for minimized tissue damage, an H-bridge architecture with a single current sink is adopted as the current driver. The active charge balancing (CB) circuitry automatically adjusts the current amplitude such that the residual voltage at the electrode-tissue interface is kept below $\pm\textit{V}_{\textit{SAFE}}$. Any residual charges are removed through passive discharging.

\begin{figure}[t]
\centering
\includegraphics[width=0.9\columnwidth]{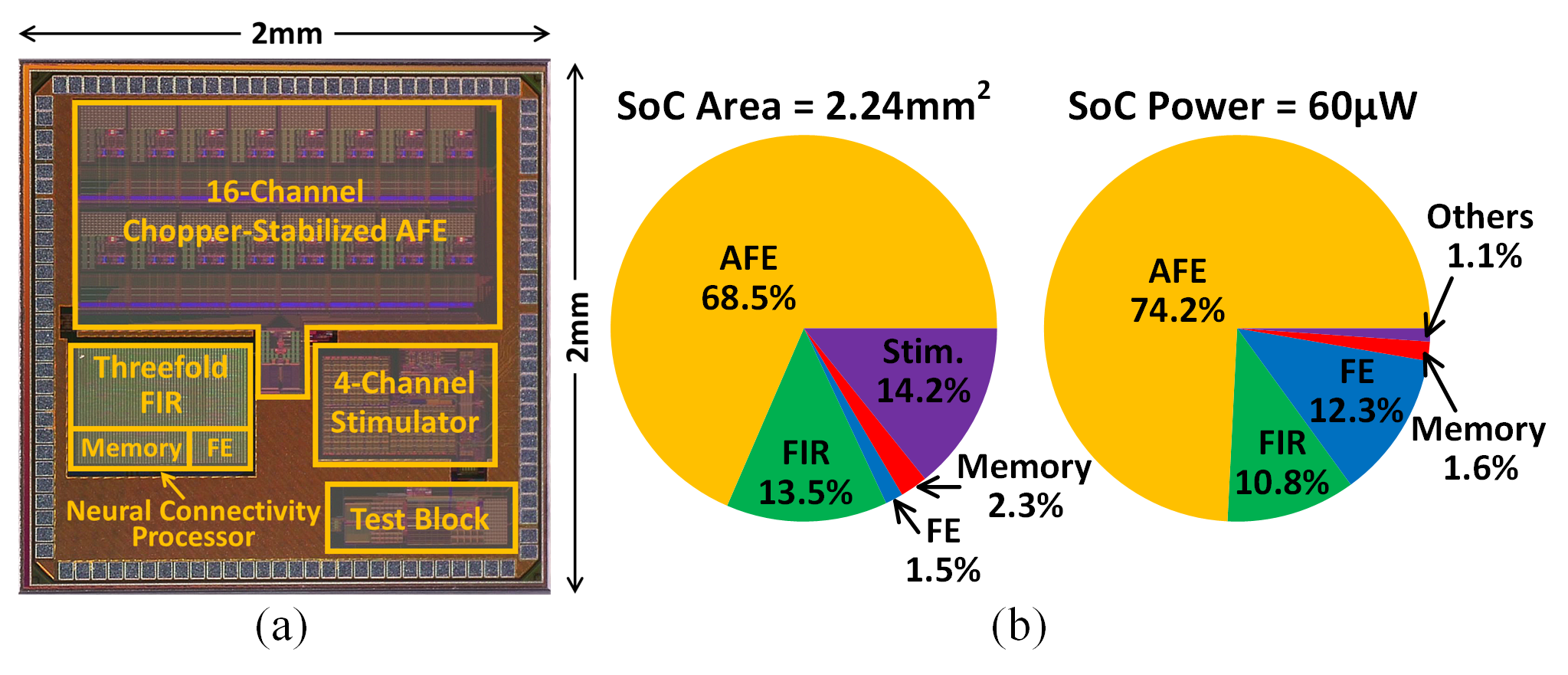}
\vspace{-3mm}
\caption{(a) Chip micrograph, and (b) SoC area and power breakdowns.}
\vspace{-3mm}
\label{f7_Chip_Breakdown}
\end{figure}

\begin{figure}[t]
\centering
\includegraphics[width=1\columnwidth]{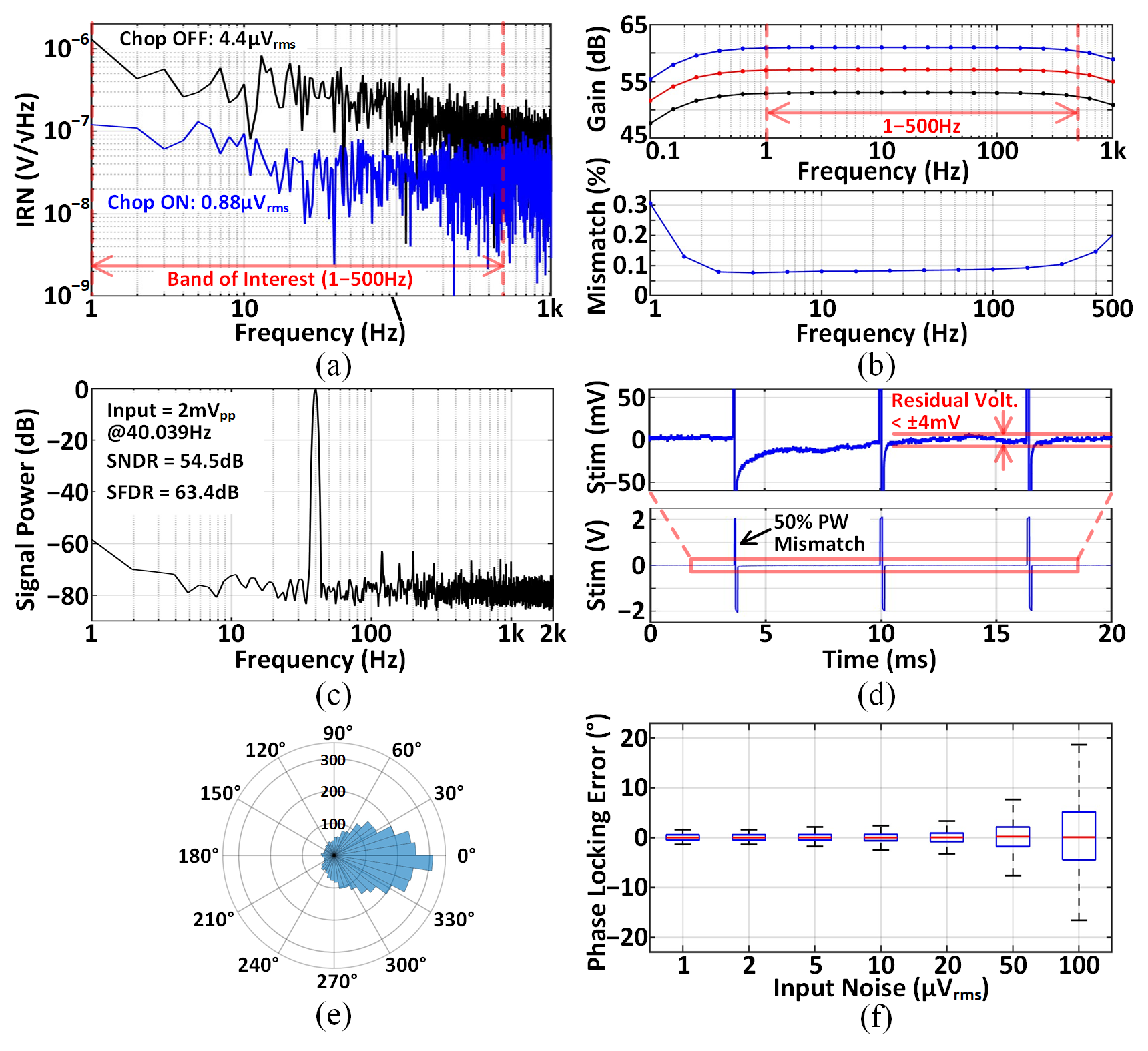}
\vspace{-7mm}
\caption{Benchtop measurement results: (a) AFE IRN performance, (b) gain programmability and 16-channel gain mismatch, (c) dynamic performance, (d) stimulator's charge balancing (5k$\Omega$+330nF load), (e) polar histogram of phase locking errors, and (f) noise tolerance in phase locking detection.}
\vspace{-3mm}
\label{f8_Benchtop}
\end{figure}

\section{Measurement Results}
A phase-locked DBS prototype was fabricated in a TSMC 65nm 1P9M CMOS process. Fig. \ref{f7_Chip_Breakdown}(a) presents a chip micrograph, and the SoC area and power breakdowns are illustrated in Fig. \ref{f7_Chip_Breakdown}(b). The SoC occupies an active area of 2.24mm\textsuperscript{2} and consumes 60$\upmu$W at 1.2V/0.85V analog/digital supply voltages.

\vspace{-3mm}
\subsection{Benchtop Measurements}
Fig. \ref{f8_Benchtop}(a) presents the input-referred noise (IRN) performance of the AFE measured at the ADC output. Chopper stabilization effectively reduced the noise from 4.4$\upmu$V\textsubscript{rms} to 0.88$\upmu$V\textsubscript{rms} in the 1--500Hz band. The gain programmability (53--61dB) is shown in Fig.~\ref{f8_Benchtop}(b). At  maximum gain setting, precise gain matching ($\sigma$/mean$<$0.1\%) across 16 channels was achieved in the mid-band. With a 2mV\textsubscript{pp}, 40.039Hz sinusoidal input, the in-band signal-to-noise and distortion ratio (SNDR) and spurious-free dynamic range (SFDR) were measured at 54.5dB and 63.4dB, respectively, as shown in Fig. \ref{f8_Benchtop}(c). Fig.~\ref{f8_Benchtop}(d) demonstrates that in the presence of an intentional 50\% pulse width mismatch, the residual voltage at the stimulator output was reduced to below $\pm$4mV by charge balancing. Fig. \ref{f8_Benchtop}(e) presents a polar histogram of phase locking errors evaluated on the pre-recorded LFPs of a Long-Evans rat. The stimulation trigger was locked to the 180$^{\circ}$ of theta-band (4--8Hz) oscillations. To account for the FIR group delay (31ms), the trigger was timed 67$^{\circ}$ ahead of the target phase, given the 6Hz center frequency (360$^{\circ}\times$31ms/166.6ms). The phase locking error is the deviation of the ground truth phase from the target at trigger instants, where the ground truth was computed using a second-order zero-phase Butterworth bandpass filter and ideal Hilbert transform in MATLAB. The histogram shows that the errors are tightly grouped around 0$^{\circ}$. Fig. \ref{f8_Benchtop}(f) demonstrates the noise tolerance in phase locking detection by the SoC. Input test signals were comprised of different levels of pink noise superimposed on a 2mV\textsubscript{pp}, 6Hz sine wave, and phase locking errors from the 180$^{\circ}$ target were measured. This suggests that the noise requirements on the AFE could be relaxed to save power without compromising phase locking accuracy.

\begin{figure}[t]
\centering
\includegraphics[width=1\columnwidth]{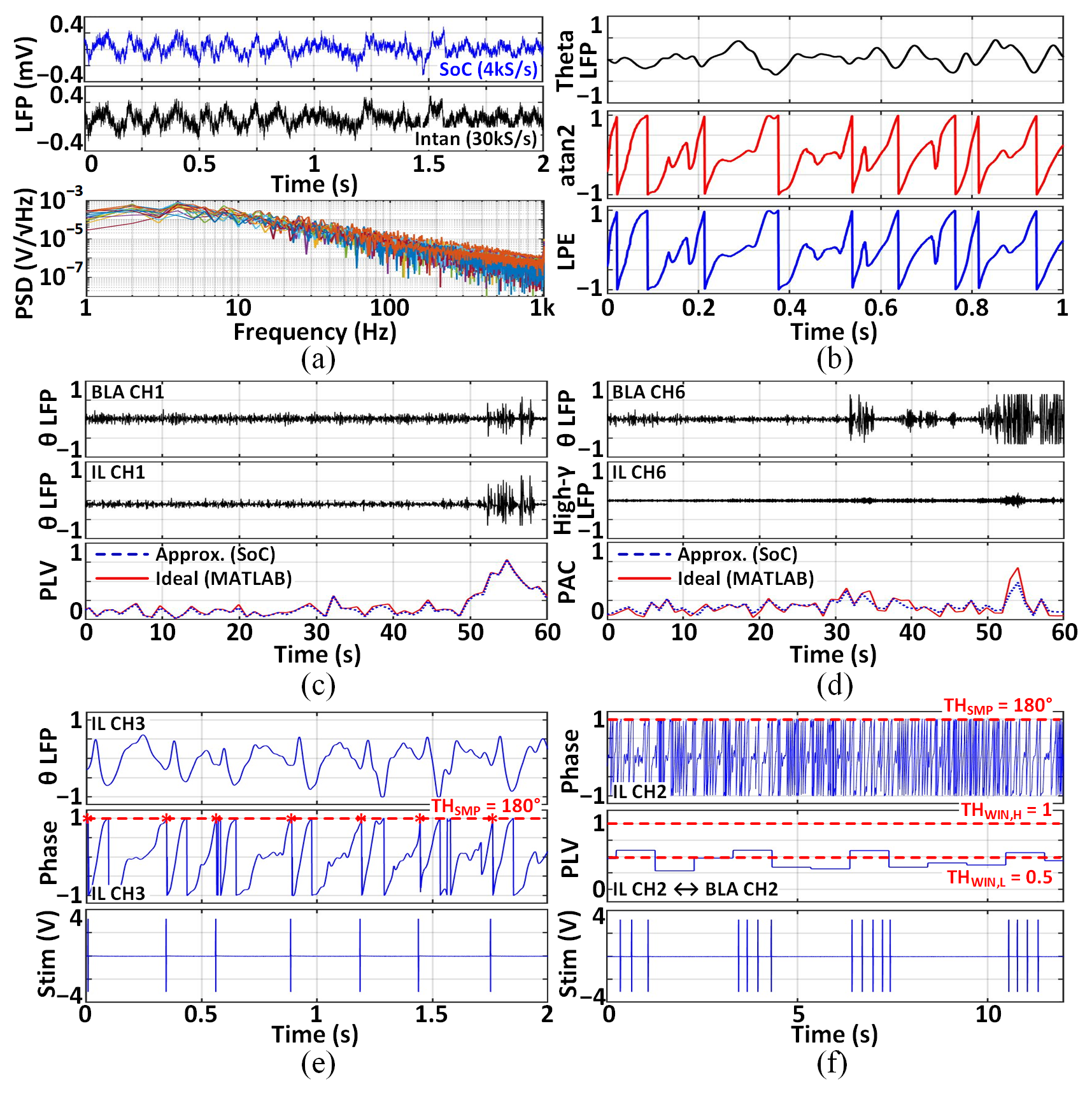}
\vspace{-7mm}
\caption{\emph{In-vivo} measurement results: (a) LFP recordings and input-referred PSD, (b) LPE phase extraction, (c) PLV, and (d) PAC extractions, (e) phase-locked stimulation, and (f) phase- and PLV-locked stimulation. The bandpass filtered LFPs, phase, PLV, and PAC are plotted in normalized units.}
\vspace{-3mm}
\label{f9_In-vivo}
\end{figure}

\vspace{-3mm}
\subsection{In-vivo Measurements}
To validate the SoC \emph{in vivo}, we implanted two arrays of 8 recording and 2 stimulation microwires into the basolateral amygdala (BLA) and infralimbic cortex (IL) of Long-Evans rats. We mainly focused on theta-band oscillations, which in this circuit (IL-BLA) are correlated with fear- and anxiety-related behavior. Fig. \ref{f9_In-vivo}(a) presents simultaneous LFP recordings by the SoC and a commercial device (Intan C3314). The 16-channel input-referred power spectral density (PSD) exhibits the expected 1/f-shaped spectra of neural oscillations. Fig. \ref{f9_In-vivo}(b) shows the LPE phase, essentially identical to the ideal one computed in MATLAB. The SoC's neural connectivity extraction is demonstrated in Figs. \ref{f9_In-vivo}(c) and (d). The approximated PLV and PAC closely track the ideal features. Figs. \ref{f9_In-vivo}(e) and (f) present closed-loop stimulation  triggered by the instantaneous phase  and a combination of cross-regional PLV and phase, respectively. Here, the maximum stimulation frequency was set to 6Hz, and the stimulator controller was designed to ignore threshold crossings due to phase wrapping. 

\begin{table}[t]
  \centering  
  \caption{Comparison with the State-of-the-Art SoCs}\vspace{-3mm}
  \includegraphics[width=1\columnwidth]{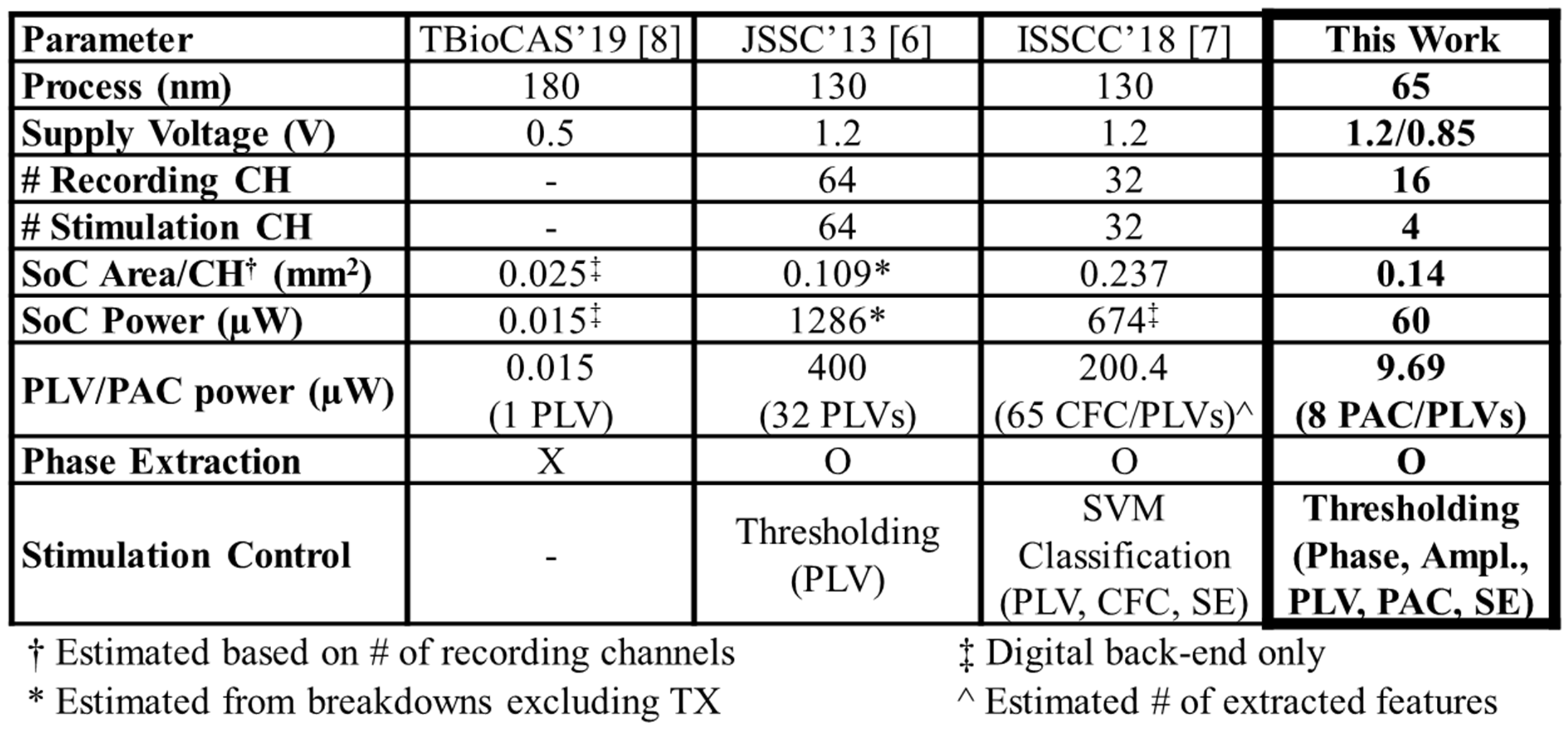}\vspace{-7mm}  
  \label{t1_Comparison}
\end{table}

Table~\ref{t1_Comparison} compares the proposed SoC with the state-of-the-art neural connectivity processors. This work provides an improved energy-accuracy trade-off in extracting the instantaneous phase and neural connectivity measures, while demonstrating phase-locked DBS for the first time. With  flexible stimulation control and various biomarkers integrated, the presented DBS scheme could serve as a new stimulation paradigm for treating a wide range of  brain disorders. 

\bibliographystyle{IEEEtran.bst}
\bibliography{cit}
\end{document}